\documentclass[12pt]{article}

\usepackage{amssymb}

\def\a{\alpha'}

\begin{document}

\hyphenation{super-sym-me-tric}

\hyphenation{cor-res-ponds}

\hyphenation{ea-si-er}

\hyphenation{li-nea-ri-zed}

\hyphenation{auto-ma-ti-cally}

\hyphenation{super-gra-vity}

\hyphenation{repre-sent}

\hyphenation{dif-fe-ren-ti-al}

\hyphenation{wri-ting}

\hyphenation{lea-ding}

\begin{titlepage}
\begin{center}

\vskip 20mm

{\Huge Type II and heterotic one loop string effective actions in
four dimensions}

\vskip 10mm

Filipe Moura

\vskip 4mm

{\em Security and Quantum Information Group - Instituto de
Telecomunica\c c\~oes\\ Instituto Superior T\'ecnico, Departamento
de Matem\'atica \\Av. Rovisco Pais, 1049-001 Lisboa, Portugal}

\vskip 4mm

{\tt fmoura@math.ist.utl.pt}

\vskip 6mm

\end{center}

\vskip .2in

\begin{center} {\bf Abstract } \end{center}
\begin{quotation}\noindent

We analyze the reduction to four dimensions of the ${\cal R}^4$
terms which are part of the ten-dimensional string effective
actions, both at tree level and one loop. We show that there are two
independent combinations of ${\cal R}^4$ present, at one loop, in
the type IIA four dimensional effective action, which means they
both have their origin in M-theory. The $d=4$ heterotic effective
action also has such terms. This contradicts the common belief that
there is only one ${\cal R}^4$ term in four-dimensional supergravity
theories, given by the square of the Bel-Robinson tensor. In pure
${\cal N}=1$ supergravity this new ${\cal R}^4$ combination cannot
be directly supersymmetrized, but we show that, when coupled to a
scalar chiral multiplet (violating the U(1) $R$-symmetry), it
emerges in the action after elimination of the auxiliary fields.
\end{quotation}

\vfill


\end{titlepage}

\eject

\tableofcontents

\section{Introduction}
\setcounter{equation}{0} \indent

String theories require higher order in $\a$ corrections to their
corresponding low energy supergravity effective actions. The
leading type II string corrections are of order $\a^3$, and
include ${\cal R}^4$ terms (the fourth power of the Riemann
tensor), both at tree level and one loop \cite{Gross:1986iv,
Grisaru:1986px}. These ${\cal R}^4$ corrections are also present
in the type I/heterotic effective actions \cite{Gross:1986mw} and
in M-theory \cite{Green:1997as}.

These string corrections to supergravity theories should obviously
be supersymmetric. Unfortunately there is still no known way to
compute these corrections in a manifestly supersymmetric way,
although important progresses have been achieved. The
supersymmetrization of these higher order string/M-theory terms
has been a topic of research for a long time \cite{Peeters:2000qj,
deRoo:1992zp}.

After compactification to four dimensions, one obtains a
supergravity theory, whose number $\mathcal{N}$ of supersymmetries
and different matter couplings depend crucially on the manifold
where the compactification is taken. Most of the times, in four
dimensions the higher order terms are studied as part of the
supergravity theories, either simple \cite{Deser:1977nt,
Moura:2001xx, Moura:2002ft} or extended \cite{Kallosh:1980fi,
Howe:1980th, Moura:2002ip, Deser:1978br}, and are therefore
considered only from a supergravity point of view. These theories
are believed to be divergent, and those are candidate
counterterms. Their possible stringy origin, as higher order terms
in string/M theory after compactification from ten/eleven
dimensions, is often neglected. One of the reasons for that
criterion is chronological: the study of the quantum properties of
four dimensional supergravity theories started several years
before superstring theories were found to be free of anomalies and
taken as the main candidates to a unified theory of all the
interactions. In higher dimensions the procedure has been
different: the low-energy limits of superstring theories are the
different ten-dimensional supergravity theories. People have
studied higher order corrections to these theories most of the
times in the context of string theory, which requires them to be
supersymmetric.

Tacitly one makes the natural assumption that, when compactified,
these higher order terms also emerge as corrections to the
corresponding four-dimensional supergravity theories. But this does
not necessarily need to be the case. The quantum behavior of these
theories is still an active topic of research, and recent works
claim that the maximal ${\cal N}=8$ theory may even be ultraviolet
finite \cite{Bern:2006kd, Green:2006gt}. If that is the case, the
${\cal N}=8$ higher order terms will not be necessary from a
supergravity point of view, although they will still appear in the
${\cal N}=8$ theory we obtain when we compactify type II
superstrings on a six-dimensional torus. All the higher order terms
considered are, from a supergravity point of view, \emph{ candidate}
counterterms; it has never been explicitly shown that they indeed
appear in the quantum effective actions with nonzero coefficients.
Even in ${\cal N}<8$ theories, it may eventually happen that some of
these counterterms are not necessary as supergravity counterterms,
but are needed as compactified string corrections.

From the known bosonic terms in the different $\a$-corrected
string effective actions in ten dimensions, one should therefore
determine precisely which terms should emerge in four dimensions
for each compactification manifold, not worrying if they are
needed in $d=4$ supergravity. This is the goal of the present
article, but here we restrict ourselves mainly to the order $\a^3$
${\cal R}^4$ terms. We will also be mainly (but not strictly)
concerned with the simplest toroidal compactifications; the reason
is that the terms one gets are "universal", i.e. they must be
present (possibly together with other moduli-dependent terms) no
matter which compactification manifold we take.

The article is organized as follows. In section 2 we review the
purely gravitational parts in the effective actions, up to order
$\a^3$, of type IIA, IIB and heterotic strings, at tree level and
one loop. In section 3 we analyze their dimensional reduction to
$d=4$. We show that there are two independent ${\cal R}^4$ terms in
the four dimensional superstring effective action, although a
classical result tells us that, of these terms, only the one which
was previously known can be directly supersymmetrized. The
supersymmetrization of the new ${\cal R}^4$ term gives rise to a new
problem, which we address in ${\mathcal N}=1$ supergravity in
section 4 by considering the coupling of the new ${\cal R}^4$ term
to a chiral multiplet in superspace.


\section{String effective actions to order $\a^3$ in $d=10$}
\setcounter{equation}{0} \indent

The Riemann tensor admits, in $d$ spacetime dimensions, the
following decomposition in terms of the Weyl tensor ${\cal
W}_{mnpq}$, the Ricci tensor ${\cal R}_{mn}$ and the Ricci scalar
${\cal R}$:
\begin{eqnarray}
{\cal R}_{mnpq}&=&{\cal W}_{mnpq}-\frac{1}{d-2} \left(g_{mp} {\cal
R}_{nq} - g_{np} {\cal R}_{mq} + g_{nq} {\cal R}_{mp} - g_{mq}
{\cal R}_{np} \right) \nonumber \\ &+&\frac{1}{(d-1)(d-2)}
\left(g_{mp} g_{nq} - g_{np} g_{mq}\right) {\cal R}.
\label{riemann}
\end{eqnarray}

As proven in \cite{Fulling:1992vm}, in $d=10$ dimensions, the critical
dimension of superstring theories, there are seven independent
real scalar polynomials made from four powers of the irreducible
components of the Weyl tensor, which we label, according to
\cite{Peeters:2000qj}, as $R_{41}, \dots, R_{46}, A_7$. These
polynomials are given by
\begin{eqnarray}
R_{41} &=& {\cal W}_{mnpq} {\cal W}^{nrqt} {\cal W}_{rstu} {\cal
W}^{smup}, \nonumber \\ R_{42} &=& {\cal W}_{mnpq} {\cal W}^{nrqt}
{\cal W}^{ms}_{\ \ \ tu} {\cal W}_{sr}^{\ \ up}, \nonumber
\\ R_{43} &=& {\cal W}_{mnpq} {\cal W}_{rs}^{\ \ pq} {\cal
W}^{mn}_{\ \ \ tu} {\cal W}^{rstu}, \nonumber
\\ R_{44} &=& {\cal W}_{mnpq} {\cal W}^{mnpq} {\cal W}_{rstu} {\cal
W}^{rstu}, \nonumber \\ R_{45} &=& {\cal W}_{mnpq} {\cal W}^{nrpq}
{\cal W}_{rstu} {\cal W}^{smtu}, \nonumber
\\ R_{46} &=& {\cal W}_{mnpq} {\cal W}_{rs}^{\
\ pq} {\cal W}^{mr}_{\ \ \ tu} {\cal W}^{nstu}, \nonumber \\ A_7
&=& {\cal W}_{mn}^{\ \ \ pq} {\cal W}^{mt}_{\ \ \ pu} {\cal
W}_{tr}^{\ \ ns} {\cal W}_{\ \ qs}^{ur}. \label{r47}
\end{eqnarray}

The superstring $\a^3$ effective actions are given in terms of two
independent bosonic terms, from which two separate superinvariants
are built \cite{Peeters:2000qj, Tseytlin:1995bi}. These terms are
given, at linear order in the NS-NS gauge field $B_{mn}$, by:
\begin{eqnarray}
I_X&=&t_8 t_8 {\cal R}^4 + \frac{1}{2} \varepsilon_{10} t_8 B
{\cal R}^4, \nonumber \\ I_Z&=&-\varepsilon_{10} \varepsilon_{10}
{\cal R}^4 +4 \varepsilon_{10} t_8 B {\cal R}^4. \label{ixiz}
\end{eqnarray}
Each $t_8$ tensor has eight free spacetime indices. It acts in
four two-index antisymmetric tensors, as defined in
\cite{Gross:1986iv, Grisaru:1986px}, where one can also find the
precise index contractions. In terms of the seven fundamental
polynomials $R_{41}, \dots, R_{46}, A_7$ from (\ref{r47}), the
purely gravitational parts of $I_X$ and $I_Z$, which we denote by
$X$ and $Z$ respectively, are given by \cite{Peeters:2000qj}:
\begin{eqnarray}
X := t_8 t_8 {\cal W}^4 &=& 192 R_{41} + 384 R_{42} + 24 R_{43} +
12 R_{44} -192 R_{45} -96 R_{46}, \nonumber \\ \frac{1}{8} Z := -
\frac{1}{8} \varepsilon_{10} \varepsilon_{10} {\cal W}^4 &=& X
+192 R_{46} - 768 A_7. \label{b14}
\end{eqnarray}

For the heterotic string two extra terms $Y_1$ and $Y_2$ appear at
order $\a^3$ at one loop level \cite{Peeters:2000qj, deRoo:1992zp,
Tseytlin:1995bi}, the pure gravitational parts of which being
given respectively by
\begin{eqnarray}
Y_1 := t_8 \left(\mbox{tr} {\cal W}^2\right)^2 &=& -4 R_{43} -2
R_{44} +16 R_{45} +8 R_{46}, \nonumber \\ Y_2 := t_8 \mbox{tr}
{\cal W}^4 &=& 8 R_{41} + 16 R_{42} -4 R_{45} -2 R_{46}.
\label{y1y2}
\end{eqnarray}
with $\mbox{tr} {\cal W}^2= {\cal W}_{mnpq} {\cal W}_{rs}^{\ \
qp}$, etc. Only three of these four invariants are independent
because, as one may see, one has the relation $X=24 Y_2 -6 Y_1.$

To be precise, let's review the form of the purely gravitational
superstring and heterotic effective actions  in the string frame
up to order $\a^3$. The perturbative terms occur at string tree
and one loop levels; there are no higher loop contributions
\cite{Green:1997as, Tseytlin:1995bi, Green:1997tv, Iengo:2002pr}.

The effective action of type IIB theory must be written, because of
its well known SL$(2,{\mathbb Z})$ invariance, as a product of a
single linear combination of order $\a^3$ invariants and an overall
function of the complexified coupling constant $\Omega= C^0 + i
e^{-\phi},$ $C^0$ being the axion. This function accounts for
perturbative (loop) and non-perturbative (D-instanton
\cite{Green:1997tv,Green:1997di}) string contributions. The
perturbative part is given in the string frame by
\begin{equation}
\left. \frac{1}{\sqrt{-g}} {\mathcal L}_{\mathrm{IIB}}
\right|_{\a^3} = -e^{-2 \phi} \a^3 \frac{\zeta(3)}{3 \times 2^{10}}
\left(I_X - \frac{1}{8} I_Z \right) - \a^3 \frac{1}{3 \times 2^{16}
\pi^5} \left(I_X - \frac{1}{8} I_Z \right). \label{2bea}
\end{equation}

Type IIA theory has exactly the same term of order $\a^3$ as type
IIB at tree level, but at one loop the sign in the coefficient of
$I_Z$ is changed when compared to type IIB:
\begin{equation}
\left. \frac{1}{\sqrt{-g}} {\mathcal L}_{\mathrm{IIA}}
\right|_{\a^3} = -e^{-2 \phi} \a^3 \frac{\zeta(3)}{3 \times 2^{10}}
\left(I_X - \frac{1}{8} I_Z \right) - \a^3 \frac{1}{3 \times 2^{16}
\pi^5} \left(I_X + \frac{1}{8} I_Z \right). \label{2aea}
\end{equation}
The reason for this sign flip is that at one string loop the
relative GSO projection between the left and right movers is
different for type IIA and type IIB, since these two theories have
different chirality properties \cite{Kiritsis:1997em,
Antoniadis:1997eg}.

Type II superstring theories only admit $\a^3$ and higher
corrections because the corresponding sigma model is two and
three-loop finite, as shown in \cite{Grisaru:1986px}: ten
dimensional ${\mathcal N}=2$ supersymmetry prevents these
corrections. Heterotic string theories have ${\mathcal N}=1$
supersymmetry in ten dimensions, which allows corrections to the
sigma model already at order $\a$, including ${\mathcal R}^2$
corrections. These corrections come both from three-graviton
scattering amplitudes and anomaly cancellation terms (the
Green-Schwarz mechanism). The effective action is then given in
the string frame, up to order $\a^3$ and neglecting the
contributions of gauge fields, by
\begin{eqnarray}
\left. \frac{1}{\sqrt{-g}} {\mathcal L}_{\mathrm{heterotic}}
\right|_{\a+ \a^3} &=& e^{-2 \phi} \left[\frac{1}{16} \a \mbox{tr}
{\cal R}^2 +\frac{1}{2^9} \a^3 Y_1 - \frac{\zeta(3)}{3 \times
2^{10}} \a^3 \left(I_X - \frac{1}{8} I_Z \right) \right] \nonumber
\\ &-& \a^3 \frac{1}{3 \times 2^{14} \pi^5} \left(Y_1+ 4 Y_2
\right). \label{hea}
\end{eqnarray}
For the type IIB theory only the combination $I_X - \frac{1}{8}
I_Z$ is present in the effective action. For the type IIA and
heterotic theories different combinations show up. The
supersymmetrization of these terms has been the object of study in
many articles \cite{Peeters:2000qj, deRoo:1992zp}, although a
complete understanding of the full supersymmetric effective
actions is still lacking. Here we are more concerned with the
number of independent superinvariants they would belong to.
Because in every theory the $I_X - \frac{1}{8} I_Z$ term includes
a transcendental factor $\zeta(3)$ (which is not shared by any
other bosonic term at the same order in $\a$), it cannot be
related to other bosonic terms by supersymmetry and requires its
own superinvariant. This way in type IIA and heterotic string
theories one then needs at least one ${\mathcal R}^4$
superinvariant for the tree level terms and another one for one
loop.

Type IIA theory comes from compactification of M-theory on ${\mathbb
S}^1$, but its tree level $\a^3$ terms vanish on the
eleven-dimensional limit, as shown in \cite{Green:1997as}. Therefore
the one-loop type IIA ${\mathcal R}^4$ term is the true
compactification of the $d=11$ ${\mathcal R}^4$ term. In M-theory,
there is only one ${\mathcal R}^4$ superinvariant. The existence of
this term was shown in \cite{Howe:2003cy}, using spinorial
cohomology, and its coefficient was fixed using anomaly cancellation
arguments. The full calculation, using pure spinor BRST cohomology,
was carried out in \cite{Anguelova:2004pg}, where it was shown that
this term is indeed unique and its coefficient can be directly
determined without using the anomaly cancellation argument.

For a more detailed review of the present knowledge of ${\mathcal
R}^4$ terms in M-theory and supergravity, including a discussion
of their supersymmetrization and related topics, see
\cite{Howe:2004pn}.


\section{String effective actions to order $\a^3$ in $d=4$}
\setcounter{equation}{0} \indent

In this section we analyze the reduction to four dimensions of the
effective actions considered in the previous section.

\subsection{${\cal R}^4$ terms in $d=4$ from $d=10$}
\indent

It is interesting to check how many independent superinvariants
one still has in four dimensions. In this case, the Weyl tensor
can still be decomposed in its self-dual and antiself-dual
parts\footnote{In the previous section, we used latin letters -
$m,n, \ldots$ - to represent ten dimensional spacetime indices.
From now on we will be only working with four dimensional
spacetime indices which, to avoid any confusion, we represent by
greek letters $\mu, \nu, \ldots$}:
\begin{equation}
{\cal W}_{\mu \nu \rho \sigma}= {\cal W}^+_{\mu \nu \rho \sigma} +
{\cal W}^-_{\mu \nu \rho \sigma}, {\cal W}^{\mp}_{\mu \nu \rho
\sigma} :=\frac{1}{2} \left({\cal W}_{\mu \nu \rho \sigma} \pm
\frac{i}{2} \varepsilon_{\mu \nu}^{\ \ \ \lambda \tau} {\cal
W}_{\lambda \tau \rho \sigma} \right), \label{wpm}
\end{equation}
which have the following properties:
\begin{equation}
{\cal W}^+_{\mu \nu \rho \sigma} {\cal W}^{- \ \rho \sigma}_{\tau
\lambda} =0, {\cal W}^\pm_{\mu \nu \rho \sigma} {\cal W}^{\pm \nu
\rho \sigma}_\tau =\frac{1}{4} g_{\mu \tau} {\cal W}^2_\pm.
\label{wprop}
\end{equation}
Besides the usual Bianchi identities, the Weyl tensor in four
dimensions obeys Schouten identities like this one:
\begin{equation}
{\cal W}^{\mu \nu}_{\ \ \ \rho \tau} {\cal W}_{\mu \nu \sigma
\lambda}= \frac{1}{4} \left(g_{\rho \sigma} g_{\tau \lambda} -
g_{\rho \lambda} g_{\tau \sigma} \right) {\cal W}^2 +2 \left({\cal
W}_{\rho \mu \nu \sigma} {\cal W}^{\ \mu \nu}_{\lambda \ \ \ \tau}
- {\cal W}_{\tau \mu \nu \sigma} {\cal W}^{\ \mu \nu}_{\lambda \ \
\ \rho} \right). \label{schouten4}
\end{equation}
Because of the given properties, the Bel-Robinson tensor, which can
be shown to be totally symmetric, is given in four dimensions by
$${\cal W}^+_{\mu \rho \nu \sigma} {\cal W}^{- \rho \ \sigma}_{\tau
\ \lambda}.$$ In the van der Warden notation, using spinorial
indices, the decomposition (\ref{wpm}) is written as
\cite{Penrose:1985jw}
\begin{equation}
{\cal W}_{A \dot A B \dot B C \dot C D \dot D}= -2
\varepsilon_{\dot A \dot B} \varepsilon_{\dot C \dot D} {\cal
W}_{ABCD}  -2 \varepsilon_{AB} \varepsilon_{CD} {\cal W}_{\dot A
\dot B \dot C \dot D} \label{wpms}
\end{equation}
with the totally symmetric ${\cal W}_{ABCD}, {\cal W}_{\dot A \dot
B \dot C \dot D}$ being given by (in the notation of
\cite{Moura:2002ft})
$${\cal W}_{ABCD}:=-\frac{1}{8} {\cal W}^+_{\mu \nu \rho \sigma}
\sigma^{\mu \nu}_{\underline{AB}} \sigma^{\rho
\sigma}_{\underline{CD}}, \, {\cal W}_{\dot A \dot B \dot C \dot
D}:=-\frac{1}{8} {\cal W}^-_{\mu \nu \rho \sigma} \sigma^{\mu
\nu}_{\underline{\dot A \dot B}} \sigma^{\rho
\sigma}_{\underline{\dot C \dot D}}.$$
Using this notation, calculations involving the Weyl tensor become
much more simplified. The Bel-Robinson tensor is simply given by
${\cal W}_{ABCD} {\cal W}_{\dot A \dot B \dot C \dot D}$.

In reference \cite{Fulling:1992vm} it is also shown that, in four
dimensions, there are only two independent real scalar polynomials
made from four powers of the Weyl tensor. Like in
\cite{Moura:2002ft}, these polynomials can be written, using the
previous notation, as
\begin{eqnarray}
{\cal W}_+^2 {\cal W}_-^2 &=& {\cal W}^{ABCD} {\cal W}_{ABCD}
{\cal W}^{\dot A \dot B \dot C \dot D} {\cal W}_{\dot A \dot B
\dot C \dot D}, \label{r441}\\ {\cal W}_+^4+{\cal W}_-^4 &=&
\left({\cal W}^{ABCD} {\cal W}_{ABCD}\right)^2 + \left({\cal
W}^{\dot A \dot B \dot C \dot D} {\cal W}_{\dot A \dot B \dot C
\dot D}\right)^2. \label{r442}
\end{eqnarray}
In particular, the seven polynomials $R_{41}, \dots, R_{46}, A_7$
from (\ref{r47}), when computed directly in four dimensions (i.e.
replacing the ten dimensional indices $m,n, \ldots$ by the four
dimensional indices $\mu, \nu, \ldots$) should be expressed in
terms of them. That is what we present in the following. For that
we wrote each polynomial in the van der Warden notation, using
(\ref{wpms}), and we used some properties of the four dimensional
Weyl tensor, like (\ref{wprop}) and (\ref{schouten4}). This way we
have shown that, \emph{in four dimensions},
\begin{eqnarray}
R_{41} &=& \frac{1}{24} {\cal W}_+^4 + \frac{1}{24} {\cal W}_-^4
-\frac{5}{8} {\cal W}_+^2 {\cal W}_-^2, \nonumber \\ R_{42} &=&
\frac{1}{12} {\cal W}_+^4 + \frac{1}{12} {\cal W}_-^4 +\frac{11}{8}
{\cal W}_+^2 {\cal W}_-^2, \nonumber \\ R_{43} &=& \frac{1}{6} {\cal
W}_+^4 + \frac{1}{6} {\cal W}_-^4 -4 {\cal W}_+^2 {\cal W}_-^2,
\nonumber \\ R_{44} &=& {\cal W}_+^4 + {\cal W}_-^4 +2 {\cal W}_+^2
{\cal W}_-^2, \nonumber \\ R_{45} &=& \frac{1}{4} {\cal W}_+^4 +
\frac{1}{4} {\cal W}_-^4 +\frac{1}{2} {\cal W}_+^2 {\cal W}_-^2,
\nonumber \\ R_{46} &=& -\frac{1}{6} {\cal W}_+^4 - \frac{1}{6}
{\cal W}_-^4 -\frac{3}{2} {\cal W}_+^2 {\cal W}_-^2, \nonumber \\
A_7 &=& -\frac{1}{24} {\cal W}_+^4 - \frac{1}{24} {\cal W}_-^4
-\frac{1}{4} {\cal W}_+^2 {\cal W}_-^2.
\end{eqnarray}

Using the definitions (\ref{b14}), we have then
\begin{eqnarray}
X &=& 24 \left({\cal W}_+^4 + {\cal W}_-^4 \right) +384 {\cal
W}_+^2 {\cal W}_-^2, \\ \frac{1}{8} Z&=& 24 \left({\cal W}_+^4 +
{\cal W}_-^4 \right) +288 {\cal W}_+^2 {\cal W}_-^2, \nonumber
\end{eqnarray}
or
\begin{eqnarray}
X - \frac{1}{8} Z &=& 96 {\cal W}_+^2 {\cal W}_-^2, \\ X +
\frac{1}{8} Z &=& 48 \left({\cal W}_+^4 + {\cal W}_-^4 \right)
+672 {\cal W}_+^2 {\cal W}_-^2.
\end{eqnarray}
$X - \frac{1}{8} Z$ is the only combination of $X$ and $Z$ which
in $d=4$ does not contain (\ref{r442}), i.e. which contains only
the square of the Bel-Robinson tensor (\ref{r441}). We find it
extremely interesting that exactly this very same combination (or,
to be precise, $I_X - \frac{1}{8} I_Z$) is, from (\ref{ixiz}), the
only one which does not depend on the ten dimensional field
$B^{mn}$ and, therefore, due to its gauge invariance, is the only
one that can appear in string theory at arbitrary loop order. This
combination is indeed present at string tree level in every
superstring theory, multiplied by a transcendental factor
$\zeta(3)$, as we have seen in the previous section.

From (\ref{y1y2}) one also derives in $d=4:$
\begin{eqnarray}
Y_1&=& 8 {\cal W}_+^2 {\cal W}_-^2, \\Y_1 + 4 Y_2 = \frac{X}{6} +
2 Y_1 &=& 80 {\cal W}_+^2 {\cal W}_-^2 + 4 \left({\cal W}_+^4 +
{\cal W}_-^4 \right).
\end{eqnarray}

As seen in the previous section, for the type IIB theory only the
combination $I_X - \frac{1}{8} I_Z$ (or ${\cal W}_+^2 {\cal
W}_-^2$ in $d=4$) is present in the effective action (\ref{2bea}).
For the type IIA and heterotic theories different combinations
show up. In these two cases, ${\cal W}_+^4 + {\cal W}_-^4$ shows
up at string one loop level in the effective actions (\ref{2aea})
and (\ref{hea}) of these theories when they are compactified to
four dimensions. At string tree level, though, for all these
theories in $d=4$ only ${\cal W}_+^2 {\cal W}_-^2$ shows up. This
fact is quite remarkable, particularly for the heterotic theory,
if we consider that the two different contributions $I_X -
\frac{1}{8} I_Z$ and $Y_1$ in (\ref{hea}) have completely
different origins.

\subsection{Moduli-independent terms in $d=4$ effective actions}
\indent

All the terms we have been considering, when taken in the Einstein
frame (which is the right frame for a supergravity analysis to be
performed), are multiplied by an adequate power of $\exp(\phi).$
To be precise, consider an arbitrary term $I_i({\cal R, M})$ in
the string frame lagrangian in $d$ dimensions. $I_i({\cal R, M})$
is a function, with conformal weight $w_i$, of any given order in
$\a$, of the Riemann tensor $\cal R$ and any other fields - gauge
fields, scalars, and also fermions - which we generically
designate by $\cal M$. To pass from the string to the Einstein
frame, we redefine the metric through a conformal transformation
involving the dilaton, given by
\begin{eqnarray} g_{mn}
&\rightarrow& \exp \left( \frac{4}{d-2} \phi \right) g_{mn},
\nonumber \\ {{\cal R}_{mn}}^{pq} &\rightarrow& \exp
\left(-\frac{4}{d-2} \phi \right) {\widetilde{{\cal R}}_{mn}}^{\ \
\ pq}, \label{rsre}
\end{eqnarray}
with ${\widetilde{{\cal R}}_{mn}}^{\ \ \ pq}={{\cal R}_{mn}}^{pq}
- {\delta_{\left[m\right.}}^{\left[p\right.} \nabla_{\left.n
\right]} \nabla^{\left.q \right]} \phi.$ The transformation above
takes $I_i({\cal R,M})$ to $e^{\frac{4}{d-2} w_i \phi} I_i({\cal
\widetilde{R}, M}).$ After considering all the dilaton couplings
and the effect of the conformal transformation on the metric
determinant factor $\sqrt{-g},$ the string frame lagrangian
\begin{equation} \label{esf} \frac{1}{2}
\sqrt{-g}\ \mbox{e}^{-2 \phi} \Big( -{\cal R} + 4 \left(
\partial^m \phi \right) \partial_m \phi + \sum_i I_i({\cal R,M}) \Big)
\end{equation} is converted into the Einstein frame lagrangian \begin{equation}
\label{eef} \frac{1}{2} \sqrt{-g} \left( -{\cal R} - \frac{4}{d-2}
\left( \partial^m \phi \right) \partial_m \phi + \sum_i
\mbox{e}^{\frac{4}{d-2} \left( 1 + w_i \right) \phi} I_i({\cal
\widetilde{R},M}) \right).
\end{equation}

We finish this section by writing, for later reference, the
effective actions (\ref{2bea}), (\ref{2aea}), (\ref{hea}) in four
dimensions, in the Einstein frame (considering only terms which
are simply powers of the Weyl tensor, without any other fields
except their couplings to the dilaton, and introducing the $d=4$
gravitational coupling constant $\kappa$):
\begin{eqnarray}
\left. \frac{\kappa^2}{\sqrt{-g}} {\mathcal L}_{\mathrm{IIB}}
\right|_{{\cal R}^4} &=& - \frac{\zeta(3)}{32} e^{-6 \phi} \a^3
{\cal W}_+^2 {\cal W}_-^2 - \frac{1}{2^{11} \pi^5} e^{-4 \phi}\a^3
{\cal W}_+^2 {\cal W}_-^2, \label{2bea4} \\ \left.
\frac{\kappa^2}{\sqrt{-g}} {\mathcal L}_{\mathrm{IIA}}
\right|_{{\cal R}^4} &=& - \frac{\zeta(3)}{32} e^{-6 \phi} \a^3
{\cal W}_+^2 {\cal W}_-^2 \nonumber \\ &-& \frac{1}{2^{12} \pi^5}
e^{-4 \phi}\a^3 \left[\left({\cal W}_+^4 + {\cal W}_-^4 \right)
+224 {\cal W}_+^2 {\cal W}_-^2 \right], \label{2aea4}
\\ \left. \frac{\kappa^2}{\sqrt{-g}} {\mathcal L}_{\mathrm{het}}
\right|_{{\cal R}^2 + {\cal R}^4} &=& -\frac{1}{16} e^{-2 \phi} \a \left({\cal
W}_+^2 + {\cal W}_-^2 \right) +\frac{1}{64} \left(1-2 \zeta(3)
\right) e^{-6 \phi} \a^3 {\cal W}_+^2 {\cal W}_-^2 \nonumber \\
&-& \frac{1}{3\times2^{12} \pi^5} e^{-4 \phi}\a^3
\left[\left({\cal W}_+^4 + {\cal W}_-^4 \right) +20 {\cal W}_+^2
{\cal W}_-^2 \right]. \label{hea4}
\end{eqnarray}
Here one must refer that these are only the moduli-independent terms
of these effective actions. Strictly speaking these are not
moduli-independent terms, since they are all multiplied by the
volume of the compactification manifold (a factor we omitted for
simplicity). But they are always present, no matter which
compactification is taken. The complete action, for every different
compactification manifold, includes many moduli-dependent terms
which we do not consider here.

A complete study of the heterotic string moduli dependent terms,
but only for $\a=0$ and for a ${\mathbb T}^6$ compactification,
can be seen in \cite{Sen:1994fa}. The tree level and one loop
contributions to the four graviton amplitude, for a
compactification on an $n$-dimensional torus ${\mathbb T}^n$ of
ten dimensional type IIA/IIB string theories, can be found in
\cite{Green:1997di}.

A detailed study of these moduli-dependent ${\cal R}^4$ terms, at
string tree level and one loop, for type IIA and IIB superstrings,
for several compactification manifolds preserving different
ammounts of supersymmetry, is available in \cite{Giusto:2004xm}.
In many cases one must consider extra contributions to the
effective action coming from string winding modes and worldsheet
instantons. For the particularly simple but illustrative case of
an ${\mathbb S}^1$ compactification (presented in detail in
\cite{Green:1997di,Giusto:2004xm}), the tree level terms for both
type IIA and IIB theories are trivial: they are simply multiplied
by the volume $2 \pi R.$ At one loop level, one gets terms
proportional to the compactification radius $R$; by applying
$T$-duality to these terms, one gets other terms proportional to
$\frac{\a}{R}.$ This way one gets the term $X+\frac{1}{8}Z$, in
$d=9$, even for type IIB effective action (in this case, only at a
higher order in $\a$). The same is true in $d=4$, for more
complicated compactification manifolds.

To conclude, for any $d=4$ compactification of heterotic or
superstring theories one has, in the respective effective action,
the two different $d=4$ ${\cal R}^4$ terms (\ref{r441}) and
(\ref{r442}), multiplied by a corresponding dilaton factor and maybe
some moduli terms. This is the most important result for the rest of
this paper. From now on we will be concerned with the
supersymmetrization of these terms.


\section{${\cal R}^4$ terms and $d=4$ supersymmetry}
\setcounter{equation}{0} \indent

Up to now, we have only been considering bosonic terms for the
effective actions, but we are interested in their full
supersymmetric completion in $d=4.$ In general each superinvariant
consists of a leading bosonic term and its supersymmetric
completion, given by a series of terms with fermions. In this work
we are particularly focusing on ${\cal R}^4$ terms.
\subsection{Some known results}
\indent

It has been known for a long time that the square of the
Bel-Robinson tensor ${\cal W}_+^2 {\cal W}_-^2$ can be made
supersymmetric, in simple \cite{Deser:1977nt, Moura:2001xx} and
extended \cite{Kallosh:1980fi, Moura:2002ip, Deser:1978br} four
dimensional supergravity. For the term ${\cal W}_+^4 + {\cal
W}_-^4$ there is a "no-go theorem", based on ${\mathcal N}=1$
chirality arguments \cite{Christensen:1979qj}: for a polynomial
$I({\cal W})$ of the Weyl tensor to be supersymmetrizable, each
one of its terms must contain equal powers of ${\cal W}^+_{\mu \nu
\rho \sigma}$ and ${\cal W}^-_{\mu \nu \rho \sigma}$. The whole
polynomial must then vanish when either ${\cal W}^+_{\mu \nu \rho
\sigma}$ or ${\cal W}^-_{\mu \nu \rho \sigma}$ do. The only
exception is ${\cal W}^2 = {\cal W}_+^2 + {\cal W}_-^2$, which in
$d=4$ is part of the Gauss-Bonnet topological term and is
automatically supersymmetric.

But the new term (\ref{r442}) is part of the heterotic and type
IIA effective actions at one loop which must be supersymmetric,
even after compactification to $d=4$. One must then find out how
this term can be made supersymmetric, circumventing the ${\mathcal
N}=1$ chirality argument from \cite{Christensen:1979qj}. That is
our main goal in this paper.

One must keep in mind the assumptions in which it was derived,
namely the preservation by the supersymmetry transformations of
$R$-symmetry which, for ${\mathcal N}=1$, corresponds to U(1) and
is equivalent to chirality. That is true for pure ${\mathcal N}=1$
supergravity, but to this theory and to most of the extended
supergravity theories (except ${\mathcal N}=8$) one may add matter
couplings and extra terms which violate U(1) $R$-symmetry and yet
can be made supersymmetric, inducing corrections to the
supersymmetry transformation laws which do not preserve U(1)
$R$-symmetry.

Since the article \cite{Christensen:1979qj} only deals with the
term (\ref{r442}) by itself, one can consider extra couplings to
it and only then try to supersymmetrize. These couplings could
eventually (but not necessarily) break U(1) $R$-symmetry. This
procedure is very natural, taking into account the scalar
couplings that multiply (\ref{r442}) in the actions (\ref{2aea4}),
(\ref{hea4}).

Considering couplings to other multiplets and breaking U(1) may be
possible in ${\mathcal N}=4$ supergravity, for ${\mathbb T}^6$
compactifications of heterotic strings, but ${\mathcal N}=1$
supergravity has the advantage of being much less restrictive than
its extended counterparts. To our purposes, the simplest and most
obvious choice of coupling is to ${\mathcal N}=1$ chiral
multiplets. That is what we do in the following subsection.

\subsection{${\cal W}_+^4 + {\cal W}_-^4$ in ${\mathcal N}=1$ matter-coupled supergravity}
\indent

The ${\mathcal N}=1$ supergravity multiplet is very simple. What
also makes this theory easier is the existence of several
different full off-shell formulations. We work in standard "old
minimal" supergravity, having as auxiliary fields a vector
$A_{A\dot A}$, a scalar $M$ and a pseudoscalar $N$, given as
$\theta=0$ components of superfields $G_{A\dot A}, R,
\overline{R}:$\footnote{The ${\mathcal N}=1$ superspace
conventions are exactly the same as in
\cite{Moura:2001xx,Moura:2002ft}.}
\begin{equation}
\left. G_{A\dot A}\right| =\frac{1}{3} A_{A\dot A},
\left. \overline{R}\right| = 4 \left( M+iN \right),
\left. R \right| = 4 \left( M-iN \right).
\end{equation}
Besides there is a chiral superfield $W_{ABC}$ and its hermitian
conjugate $W_{\dot A \dot B \dot C}$, which together at $\theta=0$
constitute the field strength of the gravitino. The Weyl tensor
shows up as the first $\theta$ term: in the notation of
(\ref{wpms}), at the linearized level,
\begin{equation}
\left.\nabla _{\underline{D}}W_{\underline{ABC}}\right|={\cal
W}_{ABCD} + \ldots \label{wabcd}
\end{equation}
${\cal W}_+^4 + {\cal W}_-^4$ is proportional to the $\theta=0$ term
of $\left( \nabla^2 W^2 \right)^2 +\mathrm{h.c.},$ which cannot
result from a superspace integration. This whole term itself is U(1)
$R$-symmetric, like $\nabla _{\underline{D}}W_{\underline{ABC}}$;
indeed, the components of the Weyl tensor are U(1) $R$-neutral,
according to the weights \cite{Moura:2002ft} $$ \nabla_A \mapsto +1,
R \mapsto +2, G_m \mapsto 0, W_{ABC} \mapsto -1.$$

This way, as expected, one needs some extra coupling to
(\ref{r442}) in order to break U(1) $R$-symmetry. We can use the
fact that there are many more matter fields with its origin in
string theory and many different matter multiplets to which one
can couple the ${\mathcal N}=1$ supergravity multiplet in order to
build superinvariants. This way we hope to find some coupling
which breaks U(1) $R$-symmetry and simultaneously supersymmetrizes
(\ref{r442}), which could result from the elimination of the
matter auxiliary fields.

Having this in mind, we consider a chiral multiplet, represented by
a chiral superfield $\mathbf{\Phi}$ (we could take several chiral
multiplets $\Phi_i$, but we restrict ourselves to one for
simplicity), and containing a scalar field $\Phi = \left.
\mathbf{\Phi} \right|$, a spin$-\frac{1}{2}$ field $\left. \nabla_A
\mathbf{\Phi} \right|$, and an auxiliary field $F=-\frac{1}{2}
\left. \nabla^2 \mathbf{\Phi} \right|$. This superfield and its
hermitian conjugate couple to ${\mathcal N}=1$ supergravity in its
simplest version through a superpotential
\begin{equation}
P\left(\mathbf{\Phi}\right)=d + a \mathbf{\Phi} + \frac{1}{2} m
\mathbf{\Phi}^2 + \frac{1}{3} g \mathbf{\Phi}^3 \label{p}
\end{equation}
and a K\"ahler potential $K\left(\mathbf{\Phi},
\overline{\mathbf{\Phi}} \right)=-\frac{3}{\kappa^2} \ln
\left(-\frac{\Omega\left(\mathbf{\Phi}, \overline{\mathbf{\Phi}}
\right)}{3} \right),$ with $\Omega\left(\mathbf{\Phi},
\overline{\mathbf{\Phi}} \right)$ given by
\begin{equation}
\Omega\left(\mathbf{\Phi}, \overline{\mathbf{\Phi}} \right)=-3+
\mathbf{\Phi} \overline{\mathbf{\Phi}} + c \mathbf{\Phi} +
\overline{c} \overline{\mathbf{\Phi}}. \label{o}
\end{equation}

In order to include the term (\ref{r442}), we take the following
effective action:
\begin{eqnarray}
{\cal L}&=&-\frac{1}{6 \kappa^2} \int E
\left[\Omega\left(\mathbf{\Phi}, \overline{\mathbf{\Phi}} \right) +
\a^3 \left(b \mathbf{\Phi} \left(\nabla^2 W^2\right)^2 +
\overline{b} \overline{\mathbf{\Phi}} \left(\overline{\nabla}^2
\overline{W}^2\right)^2 \right) \right] d^4\theta \nonumber
\\ &-&\frac{2}{\kappa^2} \left(\int \epsilon
P\left(\mathbf{\Phi}\right) d^2\theta + \mathrm{h.c.} \right)
\nonumber \\ &=& \frac{1}{4 \kappa^2} \int \epsilon \left[ \left(
\overline{\nabla}^2 +\frac{1}{3} \overline{R} \right) \left(
\Omega\left(\mathbf{\Phi}, \overline{\mathbf{\Phi}} \right) + \a^3
\left(b \mathbf{\Phi} \left(\nabla^2 W^2\right)^2 + \overline{b}
\overline{\mathbf{\Phi}} \left(\overline{\nabla}^2
\overline{W}^2\right)^2 \right) \right) \right. \nonumber \\ && \,
\, \, \, \, \, \, \, \, \, \, \, \, \, \, \, \, \, \, \, \, \, \, \,
- \left. 8 P\left(\mathbf{\Phi}\right) \right] d^2\theta +
\mathrm{h.c.}. \label{r421}
\end{eqnarray}
$E$ is the superdeterminant of the supervielbein; $\epsilon$ is the
chiral density. The $\Omega\left(\mathbf{\Phi},
\overline{\mathbf{\Phi}} \right)$ and $P\left(\mathbf{\Phi}\right)$
terms represent the most general renormalizable coupling of a chiral
multiplet to pure supergravity \cite{Cremmer:1978hn}; the extra
terms represent higher-order corrections. Of course (\ref{r421}) is
meant as an effective action and therefore does not need to be
renormalizable.

The component expansion of this action may be found using the
explicit $\theta$ expansions for $\epsilon$ and $\nabla^2 W^2$
given in \cite{Moura:2002ft}. From (\ref{wabcd}), we have
\begin{equation}
\left. \nabla^2 W^2 \right| = -2{\cal W}_+^2 + \ldots \label{d2w2}
\end{equation}

It is well known that an action of this type in pure supergravity
(without the higher-order corrections) will give rise, in
$x$-space, to a leading term given by $\frac{1}{6 \kappa^2} e
\left. \Omega \right| {\mathcal R}$ instead of the usual
$-\frac{1}{2 \kappa^2} e {\mathcal R}.$\footnote{As usual in
supergravity theories we work with the vielbein and not with the
metric. Therefore, here we write $e$, the determinant of the
vielbein, instead of $\sqrt{-g}.$} In order to remove the extra
$\Phi {\mathcal R}$ terms in $\frac{1}{6 \kappa^2} e \left. \Omega
\right| {\mathcal R}$, one takes a $\Phi,
\overline{\Phi}$-dependent conformal transformation
\cite{Cremmer:1978hn}; if one also wants to remove the higher
order $\Phi {\mathcal R}$ terms, this conformal transformation
must be $\a$-dependent. Here we are only interested in obtaining
the supersymmetrization of ${\cal W}_+^4 + {\cal W}_-^4$;
therefore we will not be concerned with the Ricci terms of any
order.

If one expands (\ref{r421}) in components, one does not directly get
(\ref{r442}), but one should look at the auxiliary field sector.
Because of the presence of the higher-derivative terms, the
auxiliary field from the original conformal supermultiplet $A_m$
also gets higher derivatives in its equation of motion, and
therefore it cannot be simply eliminated
\cite{Moura:2001xx,Moura:2002ip}. Here we only consider the much
simpler terms which include the chiral multiplet auxiliary field
$F$. Take the superfields
\begin{equation}
\mathbf{\tilde{C}}= c + \a^3 b \left(\nabla^2 W^2\right)^2,
\widetilde{\Omega}\left(\mathbf{\Phi}, \overline{\mathbf{\Phi}},
\mathbf{\tilde{C}}, \overline{\mathbf{\tilde{C}}} \right)= -3 +
\mathbf{\Phi} \overline{\mathbf{\Phi}} + \mathbf{\tilde{C}}
\mathbf{\Phi} + \overline{\mathbf{\tilde{C}}}
\overline{\mathbf{\Phi}}, \label{ctil}
\end{equation}
so that the action (\ref{r421}) becomes
\begin{equation}
\frac{1}{4 \kappa^2} \int \epsilon \left[ \left( \overline{\nabla}^2
+\frac{1}{3} \overline{R} \right)
\widetilde{\Omega}\left(\mathbf{\Phi}, \overline{\mathbf{\Phi}},
\mathbf{\tilde{C}}, \overline{\mathbf{\tilde{C}}} \right)- 8
P\left(\mathbf{\Phi}\right) \right] d^2\theta + \mathrm{h.c.}
\end{equation}
and all the $\a^3$ corrections considered in it become implicitly
included in $\widetilde{\Omega}\left(\mathbf{\Phi},
\overline{\mathbf{\Phi}}, \mathbf{\tilde{C}},
\overline{\mathbf{\tilde{C}}} \right)$ through $\mathbf{\tilde{C}},
\overline{\mathbf{\tilde{C}}}.$ We also define $\tilde{C} = \left.
\mathbf{\tilde{C}} \right|$ and the functional derivative
$P_{\mathbf{\Phi}}=\partial P/
\partial \mathbf{\Phi}.$ From now on, we will work in $x$-space and
assume there is no confusion between the superfield functionals
$\widetilde{\Omega}\left(\mathbf{\Phi}, \overline{\mathbf{\Phi}},
\mathbf{\tilde{C}}, \overline{\mathbf{\tilde{C}}} \right)$,
$P\left(\mathbf{\Phi}\right)$, $P_{\mathbf{\Phi}}$ and their
corresponding $x$-space functionals $\widetilde{\Omega}\left(\Phi,
\overline{\Phi}, \tilde{C}, \overline{\tilde{C}} \right)$,
$P\left(\Phi\right)$, $P_{\Phi}.$ The terms we are looking for are
given by \cite{Cremmer:1978hn}
\begin{eqnarray}
\kappa^2 {\cal L}_{F, \overline{F}}&=& \frac{1}{9}e
\widetilde{\Omega}\left(\Phi, \overline{\Phi}, \tilde{C},
\overline{\tilde{C}} \right) \left| M-i N
-\frac{3}{\widetilde{\Omega}\left(\Phi, \overline{\Phi}, \tilde{C},
\overline{\tilde{C}} \right)} \left(\Phi + \overline{\tilde{C}}
\right) F\right|^2 \nonumber \\ &-& e \frac{3+ \tilde{C}
\overline{\tilde{C}}}{\widetilde{\Omega}^2\left(\Phi,
\overline{\Phi}, \tilde{C}, \overline{\tilde{C}} \right)} F
\overline{F} +e \tilde{P}_{\Phi} F +e
\overline{\tilde{P}}_{\overline{\Phi}} \overline{F}. \label{ptil}
\end{eqnarray}
This equation would be exact, with $\tilde{P}_{\Phi}=P_{\Phi}$ and
$\overline{\tilde{P}}_{\overline{\Phi}} =
\overline{P}_{\overline{\Phi}}$, if we were only considering the
$\theta=0$ components of $\mathbf{\tilde{C}},
\overline{\mathbf{\tilde{C}}}$. But, of course (as it is clear from
(\ref{r421})), coupled to $F$ we will have $\nabla_{\dot A}
\left(\nabla^2 W^2\right)^2$ and $\overline{\nabla}^2 \left(\nabla^2
W^2\right)^2$ terms (and $\nabla_A \left(\overline{\nabla}^2
\overline{W}^2\right)^2$ and $\nabla^2 \left(\overline{\nabla}^2
\overline{W}^2\right)^2$ terms coupled to $\bar{F}$). These terms
will not play any role for our purpose (which is to show that there
exists a supersymmetric lagrangian which contains (\ref{r442}), and
not necessarily to compute it in full), and therefore we do not
compute them explicitly. We write them in (\ref{ptil}) because we
include them in $\tilde{P}_{\Phi}$, through the definition
(analogous for $\overline{\tilde{P}}_{\overline{\Phi}}$)
$$\tilde{P}_{\Phi}=P_{\Phi} +\left(\nabla_{\dot A}
\mathbf{\tilde{C}} + \overline{\nabla}^2 \mathbf{\tilde{C}} \,
\mathrm{terms} \right).$$

The first term in (\ref{ptil}) contains the well known term
$-\frac{1}{3} e \left(M^2 + N^2 \right)$ from "old minimal"
supergravity. Because the auxiliary fields $M, N$ belong to the
chiral compensating multiplet, their field equation should be
algebraic, despite the higher derivative corrections
\cite{Moura:2001xx,Moura:2002ip}. That calculation should still
require some effort; plus, those $M, N$ auxiliary fields should not
generate \emph{by themselves} terms which violate U(1) $R$-symmetry:
these terms should only occur through the elimination of $F,
\bar{F}.$ This is why we will only be concerned with these auxiliary
fields, which therefore can be easily eliminated through their field
equation
$$\left(\frac{\left(\overline{\Phi} + \tilde{C} \right) \left(\Phi
+ \overline{\tilde{C}} \right)}{\widetilde{\Omega} \left(\Phi,
\overline{\Phi}, \tilde{C}, \overline{\tilde{C}} \right)} - \frac{3+
\tilde{C} \overline{\tilde{C}}}{\widetilde{\Omega}^2 \left(\Phi,
\overline{\Phi}, \tilde{C}, \overline{\tilde{C}} \right)} \right) F=
- \overline{\tilde{P}}_{\overline{\Phi}} - \frac{1}{3}
\left(\overline{\Phi} + \tilde{C} \right) \left( M-i N \right).$$
Replacing $F, \bar{F}$ in ${\cal L}_{F, \overline{F}}$, one gets
\begin{equation}
\kappa^2 {\cal L}_{F, \overline{F}} =- e \frac{ \tilde{P}_{\Phi}
\overline{\tilde{P}}_{\overline{\Phi}} \widetilde{\Omega}^2
\left(\Phi, \overline{\Phi}, \tilde{C}, \overline{\tilde{C}}
\right)}{\left(\overline{\Phi} + \tilde{C} \right) \left(\Phi +
\overline{\tilde{C}} \right) \widetilde{\Omega} \left(\Phi,
\overline{\Phi}, \tilde{C}, \overline{\tilde{C}} \right) - \left(
\tilde{C} \overline{\tilde{C}} +3 \right)} + M, N\,\,
\mathrm{terms.}
\end{equation}
This is a nonlocal, nonpolynomial action. Since we take it as an
effective action, we can expand it in powers of the fields $\Phi,
\overline{\Phi}$, but also in powers of $\tilde{C},
\overline{\tilde{C}}.$ These last fields contain both the couplings
of $\mathbf{\Phi}$ to supergravity $c$ and the string parameter
$\a;$ expanding in these fields is equivalent to expanding in a
certain combination of these parameters. Here one should notice that
we are only considering up to $\a^3$ terms. If we wanted to consider
higher (than $\a^3$) order corrections, together with these we
should also have included \emph{a priori} in (\ref{r421}) the
leading higher order corrections, which should be independently
supersymmetrized. Considering solely the higher than $\a^3$ order
corrections coming directly from the elimination of (any of) the
auxiliary fields from the $\a^3$ effective action (\ref{r421}) would
be misleading. The correct expansion of (\ref{r421}) to take, in the
first place, is in $\a^3$. That is what we do in the following,
after replacing $\tilde{C}, \overline{\tilde{C}}$ by their explicit
superfield expressions given by (\ref{ctil}) and taking $\theta=0$.
We also exclude the $M, N$ contributions and the higher $\theta$
terms from $\mathbf{\tilde{C}}, \mathbf{\overline{\tilde{C}}}$ in
$\tilde{P}_{\Phi}, \overline{\tilde{P}}_{\overline{\Phi}}$, for the
reasons mentioned before: they are not significant for the term we
are looking for. The resulting lagrangian we get (which we still
call ${\cal L}_{F, \overline{F}}$ to keep its origin clear, although
it is not anymore the complete lagrangian resulting from the
elimination of $F, \bar{F}$) is
\begin{eqnarray}
\kappa^2 {\cal L}_{F, \overline{F}}&=& - e \frac{ P_{\Phi}
\overline{P}_{\overline{\Phi}} \Omega^2 \left(\Phi,
\overline{\Phi}\right)}{\left(\overline{\Phi} + c \right) \left(\Phi
+ \overline{c} \right) \Omega \left(\Phi, \overline{\Phi}\right) -
\left( c \overline{c} +3 \right)}
\\ &+& \a^3 \frac{e P_{\Phi} \overline{P}_{\overline{\Phi}} \Omega
\left(\Phi, \overline{\Phi} \right)}{\left(\left(\overline{\Phi} + c
\right) \left(\Phi + \overline{c} \right) \Omega \left(\Phi,
\overline{\Phi} \right) - \left( c \overline{c} +3 \right)\right)^2}
\left[-2 \left(
b \Phi \left(\nabla^2 W^2\right)^2\right| \right.\nonumber \\
&+& \left. \left. \overline{b} \overline{\Phi}
\left(\overline{\nabla}^2 \overline{W}^2\right)^2 \right| \right)
\left(\left(\overline{\Phi} + c \right) \left(\Phi + \overline{c}
\right) \Omega \left(\Phi, \overline{\Phi} \right) - \left( c
\overline{c} +3 \right) \right)\nonumber \\ &+& \Omega \left(\Phi,
\overline{\Phi} \right) \left( -b \overline{c} \Phi
\left.\left(\nabla^2 W^2\right)^2\right| - \overline{b} c
\overline{\Phi} \left.\left(\overline{\nabla}^2
\overline{W}^2\right)^2 \right| \right.\nonumber \\
&+& \left(\overline{\Phi} + c \right) \left(\Phi + \overline{c}
\right) \left(b \Phi \left.\left(\nabla^2 W^2\right)^2\right|  +
\left. \overline{b} \overline{\Phi} \left(\overline{\nabla}^2
\overline{W}^2\right)^2 \right| \right)\nonumber \\ &+& \left.
\left. \Omega \left(\Phi, \overline{\Phi} \right) \left( b \left(
\overline{c} +\Phi \right)\left.\left(\nabla^2 W^2\right)^2\right| +
\overline{b} \left(c+ \overline{\Phi} \right)
\left.\left(\overline{\nabla}^2 \overline{W}^2\right)^2 \right|
\right) \right) \right] + \ldots \nonumber
\end{eqnarray}
If we look at the last line of the previous equation, we can already
identify the term we are looking for. This is still a nonlocal,
nonpolynomial action, which we expand now in powers of the fields
$\Phi, \overline{\Phi}$ coming from the denominators and the
$P_{\Phi} \overline{P}_{\overline{\Phi}}$ factors. We obtain
\begin{eqnarray}
\kappa^2 {\cal L}_{F, \overline{F}}&=& -15 e \frac{ \left(3 + c
\overline{c}\right)}{\left(3 + 4 c \overline{c}\right)^2} \left(m
\overline{a} \Phi + \overline{m} a \overline{\Phi} \right) \left(c
\Phi + \overline{c} \overline{\Phi} \right) \nonumber \\ &+& e
\frac{2 c^3 \overline{c}^3 + 60 c^2 \overline{c}^2 + 117 c
\overline{c}-135}{\left(3 + 4 c \overline{c}\right)^3} a
\overline{a} \Phi \overline{\Phi} - 36 \a^3 e \left( b
\overline{c} \left(\nabla^2 W^2\right)^2\right| \nonumber \\ &+&
\overline{b} c \left. \left. \left(\overline{\nabla}^2
\overline{W}^2\right)^2 \right| \right) \frac{a \overline{a} + m
\overline{a} \Phi + \overline{m} a \overline{\Phi} + g
\overline{a} \Phi^2 + \overline{g} a \overline{\Phi}^2 + m
\overline{m} \Phi \overline{\Phi} }{\left(3 + 4 c
\overline{c}\right)^2} \nonumber
\\ &-& 3 \a^3 a\overline{a} \frac{74 c^2 \overline{c}^2 + 192 c
\overline{c}-657}{\left(3 + 4 c \overline{c}\right)^4} \Phi
\overline{\Phi} \left( b \overline{c} \left(\nabla^2
W^2\right)^2\right| + \overline{b} c \left. \left.
\left(\overline{\nabla}^2 \overline{W}^2\right)^2 \right| \right)
\nonumber \\ &+& 15 \a^3 e \frac{a \overline{a} + m \overline{a}
\Phi + \overline{m} a \overline{\Phi}}{\left(3 + 4 c
\overline{c}\right)^3} \left[ \left( \overline{c}^2 \left(21 + 4 c
\overline{c}\right) \overline{\Phi} + \left(-9 + 6 c
\overline{c}\right) \Phi \right) b \left(\nabla^2
W^2\right)^2\right| \nonumber \\ &+& \left. \left( c^2 \left(21 +
4 c \overline{c}\right) \Phi + \left(-9 + 6 c \overline{c}\right)
\overline{\Phi} \right) \overline{b} \left.
\left(\overline{\nabla}^2 \overline{W}\right)^2\right| \right] +
\ldots \label{r42s}
\end{eqnarray}
This way we are able to supersymmetrize ${\cal W}_+^4 + {\cal
W}_-^4$, although we had to introduce a coupling to a chiral
multiplet. These multiplets show up after $d=4$ compactifications
of superstring and heterotic theories and truncation to ${\mathcal
N}=1$ supergravity \cite{Cecotti:1987nw}. Since from (\ref{d2w2})
the factor in front of ${\cal W}_+^4$ (resp. ${\cal W}_-^4$) in
(\ref{r42s}) is given by $\frac{72 b \overline{c} a \overline{a}}{\left(3 + 4 c
\overline{c}\right)^2}$ (resp. $\frac{72 \overline{b} c a
\overline{a}}{\left(3 + 4 c \overline{c}\right)^2}$), for this
supersymmetrization to be effective, the factors $a$ from
$P\left(\Phi\right)$ in (\ref{p}) and $c$ from $\Omega \left(\Phi,
\overline{\Phi}\right)$ in (\ref{o}) (and of course $b$ from
(\ref{r421})) must be nonzero.

The action (\ref{r42s}) includes the ${\mathcal N}=1$
supersymmetrization of ${\cal W}_+^4 + {\cal W}_-^4$, but without
any coupling to a scalar field or only with couplings to powers of
the scalar field from the chiral multiplet, which may be seen as
compactification moduli. But, as one can see from (\ref{2aea4}),
(\ref{hea4}), this term should be coupled to powers of the
dilaton. It is well known \cite{Cecotti:1987nw} that in ${\mathcal
N}=1$ supergravity the dilaton is part of a linear multiplet,
together with an antisymmetric tensor field and a Majorana
fermion. One must then work out the coupling to supergravity of
the linear and chiral multiplets. As usual one starts from
conformal supergravity and obtain Poincar\'e supergravity by
coupling to compensator multiplets which break superconformal
invariance through a gauge fixing condition. When there are only
chiral multiplets coupled to supergravity \cite{Cremmer:1978hn},
this gauge fixing condition can be generically solved, so that a
lagrangian has been found for an arbitrary coupling of the chiral
multiplets. In the presence of a linear multiplet, there is no
such a generic solution of the gauge fixing condition, which must
be solved case by case. Therefore, there is no generic lagrangian
for the coupling of supergravity to linear multiplets. We shall
not consider this problem here, like we did not in
\cite{Moura:2001xx, Moura:2002ft}. In both cases we were only
interested in studying the ${\mathcal N}=1$ supersymmetrization of
the two different $d=4$ ${\cal R}^4$ terms. The coupling of a
linear multiplet to these terms can be determined following the
procedure in \cite{Derendinger:1994gx}.


\subsection{${\cal W}_+^4 + {\cal W}_-^4$ in extended supergravity}
\indent

${\cal W}_+^4 + {\cal W}_-^4$ must also
arise in extended $d=4$ supergravity theories, for the reasons we saw, but the
"no-go" result of (\cite{Christensen:1979qj}) should remain valid, since
it was obtained for ${\mathcal N}=1$ supergravity, which can always be
obtained by truncating any extended theory. For extended supergravities,
the chirality argument should be replaced by preservation by supergravity
transformations of U(1), which is a part of $R$-symmetry.

${\mathcal N}=2$ supersymmetrization of ${\cal W}_+^4 + {\cal W}_-^4$
should work in a way similar to what we saw for ${\mathcal N}=1$.
${\mathcal N}=2$ chiral superfields must be Lorentz and SU(2) scalars
but they can have an arbitrary U(1) weight, which allows supersymmetric
U(1) breaking couplings.

A similar result should be more difficult to implement for
${\mathcal N} \geq 3$, because there are no generic chiral superfields.
Still, there are other multiplets than the Weyl, which one can consider
in order to couple to ${\cal W}_+^4 + {\cal W}_-^4$ and allow for its
supersymmetrization. The only exception is ${\mathcal N}=8$ supergravity,
which only allows for the Weyl multiplet. ${\mathcal N}=8$
supersymmetrization of ${\cal W}_+^4 + {\cal W}_-^4$ should therefore
be a very difficult problem, which we expect to study in a future work.

Related to this is the issue of possible finiteness of ${\mathcal
N}=8$ supergravity, which has been a recent topic of research. A
linearized three-loop candidate (the square of the Bel-Robinson
tensor) has been presented in \cite{Kallosh:1980fi}. But recent
works \cite{Bern:2006kd} show that there is no three-loop divergence
(which includes the two ${\cal R}^4$ terms). Power-counting analysis
from unitarity cutting-rule techniques predicted the lowest
counterterm to appear at least at five loops \cite{Bern:1998ug}. An
improved analysis based on harmonic superspace power-counting
improved this lower limit to six loops \cite{Howe:2002ui}. In
\cite{Howe:1980th} a seven loop counterterm was proposed, but in
\cite{Green:2006gt} it is proposed from string perturbation theory
arguments that the four graviton amplitude may be eight-loop finite.
The claim in \cite{Bern:2006kd} is even stronger: ${\mathcal N}=8$
supergravity may have the same degree of divergence as ${\mathcal
N}=4$ super-Yang-Mills theory and may therefore be ultraviolet
finite. But no definitive calculations have been made yet to prove
that claim; up to now, there is no firmly established example of a
counterterm which does not arise in the effective actions but would
be allowed by superspace non-renormalization theorems.

Because of all these open problems, we believe that higher order
terms in ${\mathcal N}=8$ supergravity definitely deserve further
study.


\section{Conclusions}\setcounter{equation}{0}
\indent

In this paper, we analyzed in detail the reduction to four
dimensions of the purely gravitational higher-derivative terms in
the string effective actions, up to order $\a^3$, for heterotic
and type IIA/IIB superstrings. From this analysis we have shown
that in the four dimensional heterotic and type IIA string
effective actions there must exist, besides the usual square of
the Bel-Robinson tensor ${\cal W}_+^2 {\cal W}_-^2$, a new ${\cal
R}^4$ term given in terms of the Weyl tensor by ${\cal W}_+^4 +
{\cal W}_-^4$. This new term results from the dimensional
reduction of the order $\a^3$ effective actions, at one string
loop, of these theories. By requiring four dimensional
supersymmetry, this term must be, like any other, part of some
superinvariant, but it had been shown, under some assumptions
(conservation of chirality), that such a superinvariant could not
exist by itself in pure ${\mathcal N}=1$ supergravity. But, by
taking a specific (chirality-breaking) coupling of this term to a
chiral multiplet in ${\mathcal N}=1$ supergravity, we were indeed
able to obtain the desired superinvariant. The ${\cal W}_+^4 +
{\cal W}_-^4$ term appeared after elimination of its auxiliary
fields, by itself, without any couplings to the chiral multiplet
fields.

To summarize, we have demonstrated the existence of a new ${\cal
R}^4$ superinvariant in $d=4$ supergravity, a result that many
people would find unexpected. The supersymmetrization of this new
${\cal R}^4$ term in extended supergravity remains an open
problem, but we found it in ${\mathcal N}=1$ supergravity. As we
concluded from our analysis of the dimensional reduction of order
$\a^3$ gravitational effective actions, this new ${\cal R}^4$ term
has its origin in the dimensional reduction of the corresponding
term in M-theory, a theory of which there is still a lot to be
understood. We believe therefore that the complete study of this
term and its supersymmetrization deserves further attention in the
future.


\section*{Acknowledgments} \indent

I wish to thank Pierre Vanhove for very important discussions,
suggestions and comments on the manuscript. I also wish to thank
Paul Howe for very useful correspondence and Martin Ro\v cek for
nice suggestions and for having persuaded me to consider the
${\mathcal N}=1$ case. It is a pleasure to acknowledge the
excellent hospitality of the Service de Physique Th\'eorique of
CEA/Saclay in Orme des Merisiers, France, where some parts of this
work were completed.

This work has been supported by Funda\c c\~ao para a Ci\^encia e a
Tecnologia through fellowship BPD/14064/2003 and Centro de L\'ogica
e Computa\c c\~ao (CLC).


\end{document}